# A Seamless Integration of Association Rule Mining with Database Systems

*Raj P. Gopalan*     *Tariq Nuruddin*     *Yudho Giri Sucahyo*

School of Computing, Curtin University of Technology
Bentley, Western Australia 6102
{raj, nuruddin, sucahyoy}@computing.edu.au

**Abstract**

The need for Knowledge and Data Discovery Management Systems (KDDMS) that support ad hoc data mining queries has been long recognized. A significant amount of research has gone into building tightly coupled systems that integrate association rule mining with database systems. In this paper, we describe a seamless integration scheme for database queries and association rule discovery using a common query optimizer for both. Query trees of expressions in an extended algebra are used for internal representation in the optimizer. The algebraic representation is flexible enough to deal with constrained association rule queries and other variations of association rule specifications. We propose modularization to simplify the query tree for complex tasks in data mining. It paves the way for making use of existing algorithms for constructing query plans in the optimization process. How the integration scheme we present will facilitate greater user control over the data mining process is also discussed. The work described in this paper forms part of a larger project for fully integrating data mining with database management.

## 1. Introduction

Major enterprises have collected large amounts of data for years from their day-to-day operations. As the volume of accumulated data has become huge, traditional database management systems are no longer adequate to support all the potential uses of such data for planning, analysis and decision-making. Therefore research has turned to the development of Data Warehousing and On-line Analytical Processing (OLAP) tools to support applications such as Decision Support and Executive Information Systems (Chaudhuri & Dayal 1997). Another significant need is for tools that automate the discovery of interesting patterns present in the data. Data Mining also known as Knowledge and Data Discovery in Databases was developed to deal with this problem. Research in Data Mining covers a number of areas including Association Rules, Sequential Patterns, Classification, and Clustering. Among these, Association Rules has received the most attention from data mining researchers so far.

Association Rules are used to identify relationships among sets of items. They are useful in several domains such as the analysis of market basket transactions in retail stores where this information is useful to increase the effectiveness of advertising, marketing, inventory control, and stock location on the shop floor. Many algorithms for mining association rules have been developed that include Apriori (Agrawal, Imielinski & Swami 1993) (Agrawal & Srikant 1994), Hashing (Park, Chen & Yu 1995), OCD (Mannila, Toivonen & Verkamo 1994), Partition (Savasere, Omiecinski & Navathe 1995), Dynamic Itemset Counting (Brin et al. 1997), SETM (Houtsma & Swami 1995), and CARMA (Hidber 1999). Many of these algorithms either use simple file systems or transfer the data from the database, process them and send them back to the database (loosely-coupled) and therefore do not take full advantage of facilities provided by existing Database Management Systems (DBMS).

Imielinski and Mannila suggested that data mining needed new concepts and methods especially for processing data mining queries (Imielinski & Mannila 1996). They foresaw the need to develop a second-generation of data mining systems for managing data mining applications similar to DBMSs that manage business applications. Much of the subsequent research in this area has focused on developing tightly coupled systems that make use of DBMS features for data mining.

A number of tightly coupled integration schemes between data mining and database systems have been reported. Agrawal and Shim described the integration of data mining with IBM DB2/CS RDBMS using User Defined Function (UDF) (Agrawal & Shim 1996). Exploration of four architectural alternatives (CacheMine, Stored Procedure, User Defined Function and Hybrid SQL-OR) and comparisons between them on performance, storage overhead, parallelization, development ease, maintenance ease and inter-operability were presented in (Sarawagi, Thomas & Agrawal 1998). Meo et al. described a modular approach to integrate their MINE RULE operator with a Relational Database System (Meo, Psaila & Ceri 1998b). Another architecture to integrate data mining with RDBMS based on the concept of query flocks was presented in (Nestorov & Tsur 1999). Lakshmanan et al. developed a query optimizer for Constrained Frequent Set Queries (CFQs) that produces optimized computation strategies for two general forms of constraints (Lakshmanan et al. 1999).

Several researchers have proposed query language extensions to specify user requests for the discovery of association rules. Imielinski et al. introduced the MINE operator as a query language primitive for database mining that can be embedded in a general programming language in order to provide an Application Programming Interface (Imielinski, Virmani & Abdulghani 1996). Han et al. proposed DMQL as another query language for data mining on relational databases (Han et al. 1996). Meo et al. described MINE RULE as an extension to SQL, complete with examples dealing with several variations to the association rule specifications (Meo, Psaila & Ceri 1998a). However, these proposals focus on language specifications rather than algorithms or techniques for optimizing the queries.

Chaudhuri suggested that data mining should take advantage of the primitives for data retrieval provided by database systems (Chaudhuri 1998a). However, the operators used for implementing SQL are not sufficient to support data mining applications (Imielinski & Mannila 1996). Meo et al. gave the semantics of their MINE RULE operator using a set of nested relational algebra operators (Meo, Psaila & Ceri 1998a). Probably because their objective was only to describe the semantics of MINE RULE, the expressions were far too complex for internal representation of queries or for performing optimization.

In this paper, we describe the general framework for an optimizer that can deal with both database and data mining queries. A common set of algebraic operators is used for internal representation in the optimizer of both classes of queries. Therefore, our query optimization scheme seamlessly integrates data mining with database systems. We present the operators using an extended relational algebra notation. The mining query is represented as an algebraic query tree that can be logically transformed using rules, and assigned algorithms to generate alternative execution plans from which an optimal plan can be chosen based on cost estimates.

The main contributions of this paper are as follows:
1. We present a scheme for seamless integration of data mining queries with database systems. It is part of a larger project for integrating data mining and database management.
2. The framework of a common query optimizer for data mining and database systems is described. It forms the core of the integrated system for data mining and database processing.
3. Query trees based on an extended nested relational algebra are used for internal representation of association rule queries in the optimizer. The expressiveness of the algebra is shown by suitable examples of typical queries including a constrained association query.
4. Since the algebra we use is a superset of a nested relational algebra, it provides a common representation scheme for both database and data mining queries.
5. We propose modularization of some common sequences of algebra operations to simplify the query tree for data mining. The concept of modules also facilitates the generation of alternative query plans for query optimization, using existing algorithms in the literature as low-level procedures.
6. The users can get more control over the mining process by allowing them to stop or pause the computations along particular branches of the query tree. We explain how this is achieved by providing optional break points between specific nodes of the query tree.
7. As far as we know, no previous paper has discussed the use of a common algebra for integrating data mining with database systems in a single query optimizer as we do. Though the algebraic operators we use are not new and the query optimization scheme is the same as for relational databases, what we describe is distinct from previous approaches for integration. It is also much simpler than most other proposals.

The structure of the rest of this paper is as follows: In Section 2, we provide the definition of terms and describe briefly the operators of the algebra. The algebraic representation of a query for discovering classical association rules is presented with an example in Section 3. The concept of modules for simplifying the query tree is also discussed in that Section. We discuss the user control of the data mining process and present the query tree for a typical constrained association query (CAQ) in Section 4. The framework of an optimizer for data mining systems and the optimization of data mining queries are discussed in Section 5. Section 6 contains conclusions and pointers for further work.

## 2. Preliminaries

In this section, we define the terms used for describing association rule mining and introduce the operators of our extended algebra. These are adapted from the existing literature.

### 2.1 Definition of Terms

We give the basic terms needed for describing association rules using the formalism of (Agrawal, Imielinski & Swami 1993). Let $I=\{i_1, i_2, \ldots, i_n\}$ be a set of items, and $D$ be a set of transactions, where a transaction $T$ is a subset of $I$ ($T \subseteq I$). Each transaction is identified by a *TID*. An association rule is an expression of the form $X \Rightarrow Y$, where $X \subset I$, $Y \subset I$ and $X \cap Y = \emptyset$. Note that each of X and Y is a set of one or more items and the quantity of each item is not considered. *X* is referred to as the *body* of the rule and *Y* as the *head*. An example of association rule is the statement that 80% of transactions that purchase A also purchase B and 10% of all transactions contain both of them. Here, 10% is the *support* of the itemset {A, B} and 80% is the *confidence* of the rule $A \Rightarrow B$. An itemset is called a *large itemset* or *frequent itemset* if its *support* is greater than a *support threshold* specified by the user, otherwise the itemset is *small* or *not frequent*. An association with the *confidence* greater than a *confidence threshold* is considered as a valid association rule.

| tid | cust | item | date | price | qty |
|---|---|---|---|---|---|
| 1 | C1 | Joystick (J) | 25/06/2001 | 30 | 2 |
| 1 | C1 | Hannibal (H) | 25/06/2001 | 40 | 1 |
| 2 | C2 | CD-RW Driver (C) | 26/06/2001 | 250 | 2 |
| 2 | C2 | Batman Returns (B) | 26/06/2001 | 35 | 1 |
| 2 | C2 | Joystick | 26/06/2001 | 30 | 1 |
| 3 | C3 | Scanner (S) | 27/06/2001 | 140 | 5 |
| 3 | C3 | Joystick | 27/06/2001 | 30 | 3 |
| 4 | C4 | CD-RW Driver | 27/06/2001 | 250 | 1 |
| 4 | C4 | Batman Returns | 27/06/2001 | 35 | 1 |
| 4 | C4 | Joystick | 27/06/2001 | 30 | 10 |

(a) The Purchase Table for Department Store X

| BODY (BD) | HEAD (HD) | sup | conf |
|---|---|---|---|
| {Batman Returns} | {CD-RW Driver} | 0.5 | 1 |
| {Batman Returns} | {Joystick} | 0.5 | 1 |
| {Batman Returns} | {CD-RW Driver, Joystick} | 0.5 | 1 |
| {CD-RW Driver} | {Batman Returns} | 0.5 | 1 |
| {CD-RW Driver} | {Joystick} | 0.5 | 1 |
| {CD-RW Driver} | {Batman Returns, Joystick} | 0.5 | 1 |
| {Batman Returns, CD-RW Driver} | {Joystick} | 0.5 | 1 |
| {Batman Returns, Joystick } | {CD-RW Driver} | 0.5 | 1 |
| { CD-RW Driver, Joystick} | {Batman Returns} | 0.5 | 1 |

(b) Desired Output

**Figure 1**: An Example of Association Rules

Suppose a user wants to extract association rules that fulfill 30% minimum support and 60% minimum confidence from the Purchase Table of Figure 1a. The desired output of association rules for this sample data is shown in Figure 1b.

## 2.2 Operators of the Algebra

To build a query optimizer for data mining, we need a suitable internal representation scheme for queries. In this paper, we use a nested relational algebra for expressing data mining queries. Nested relations were originally proposed as an alternative to the flat or first normal form relations to overcome some of the limitations of Codd's relational model (Colby 1989). In nested relations, the attributes can have non-atomic values that can themselves be relations. Algebras for nested relations can be used to manipulate sets of tuples at all levels of nesting. The concepts of nested relations have now become an integral part of object-relational database systems. Nested algebra operations have been integrated into most of the current object-relational query languages. In this section, we describe a set of operators that would be used in our optimizer framework. In the subsequent sections, we give expressions of this algebra for data mining.

The operators we need for expressing association rule discovery are SELECT, POWERSET, NEST, UNNEST, PROJECT, GROUPING, CARDINALITY, and JOIN. Due to space limitations, each operator is described only briefly, but formal definitions of all operators are available in an extended version of the paper (Gopalan, Nuruddin & Sucahyo 2001). Examples of using these operators can be found in Figures 2 and 3. Please note that the usual set operators of UNION, DIFFERENCE, INTERSECTION and CARTESIAN PRODUCT are omitted here, though they are part of the algebra.

1. **SELECT, $\sigma$**: It returns a collection of tuples in a relation, *R* that satisfy a given condition. In general, the select condition is a boolean combination (i.e., an expression using the connectives $\wedge$ and $\vee$) of terms that have the form (attr *op* constant) or (attr1 *op* attr2) where *op* could be a scalar comparison ($<, \leq, =, \neq, \geq, >$), a set comparison ($\subset, \subseteq, \not\subset, =, \supset, \supseteq$), or a set membership operator ($\in, \notin$).

2. **POWERSET, $\wp$**: It generates a powerset of values in a specified attribute. For a specified set valued attribute containing n values in a tuple, the POWERSET operator generates a tuple in the output relation, by replacing the input set value by its $2^n - 1$ subsets ($\varnothing$ is not included).

3. **NEST, $\Gamma$**: This operator groups tuples together based on common values of specified nesting attributes. The resulting relation will contain exactly one tuple for each combination of values of the nesting attributes.

4. **UNNEST, $\eta$**: It undoes the effect of NEST although it is not an inverse operation. It restructures the set of tuples and flattens out the relation so that each set member (tuple) may be examined individually.

5. **PROJECT, $\pi$**: The result of this operator can be made up of existing attributes in the input relation as well as new attributes specified using the lambda abstraction. The PROJECT operator can be applied on attributes at all levels of a nested relation without restructuring. The general form of PROJECT is given as follows:

$$\pi(\langle (A_1, f_1), (A_2, f_2), ...,(A_n, f_n)\rangle) R = \{ \langle A_1:f_1(r), A_2:f_2(r), ..., A_n:f_n(r)\rangle \mid r \in R \},$$

where R is a nested relation and $A_i$'s are unique attribute names for $1 \leq i \leq n$. Each $f_i$ is a function which takes a tuple of relation R as input and returns a value for the corresponding attribute, $A_i$.

6. **GROUPING, $\Im$**: This operator groups tuples based on the grouping attributes and computes aggregate functions (average, count, max, min, sum, etc.) on other specified attributes. The syntax of this operator is of the form,

$$_{(\text{grouping attributes})}\Im_{(\text{function attribute list})} R.$$

It partitions the relation *R* by distinct values of the grouping attributes. The schema of the result contains the grouping attributes (attribute-list in front of $\Im$) and a new attribute for each element in the function list (a list of <function><attribute> pairs after $\Im$). For each pair of function name and attribute name in the function list, the attribute name in the result relation has the form, *functionname_attributename.*

7. **CARDINALITY, ς**: This operator is used to count the number of values of a set valued attribute in each tuple of a relation.

8. **JOIN, ⋈**: This operator combines every pair of related tuples of two specified relations that satisfy the join condition into a single tuple of the result relation. A JOIN condition is a conjunction of comparisons, each of which can be a scalar comparison ($<, \leq, =, \neq, \geq, >$) or a set comparison ($\subset, \subseteq, \not\subset, =, \supset, \supseteq$) between attributes of the two relations.

## 3. Query Trees for Mining Association Rules

In Section 3.1, we present a query tree for finding all association rules for a set of transactions. It corresponds to the conventional mode of association rule mining, where the discovery system functions as a black box taking the transactions as input to produce all the rules that satisfy the specified minimum support and confidence levels. Unlike the conventional algorithms, however, the query trees facilitate user control of the mining process and optimization as discussed later in Sections 4 and 5.

The repetitive sequences of algebraic operators in query trees can be grouped into modules. This is discussed in Section 3.2.

### 3.1 Query Tree to find all Association Rules

As mentioned in (Agrawal, Imielinski & Swami 1993), the mining of association rules can be composed of two phases:
1. Find the frequent itemsets or large itemsets,
2. Use the large itemsets to generate association rules (by filtering the large itemsets that meet the *confidence threshold*).

The algebraic expressions for discovering association rules are represented as a query tree in Figure 2. The query tree is divided into two phases. The first phase identifies all the frequent itemsets. The association rules are extracted in the second phase.

The first phase consists of Steps 1 to 7 in the query tree of Figure 2. The attributes tid and item are projected from the input relation R in Step 1. Then the tuples are nested on tid in Step 2 and the powerset of items for each tid is generated in Step 3. The output of Step 3 is unnested on the itemset attribute in Step 4 so that the number of occurrences of each itemset can be counted in Step 5. In Step 6, the itemsets that would fall below the minimum support are pruned and the support value computed in Step 7. As mentioned in the previous section, the cost of powerset operation can be very high so we need to combine powerset generation with pruning of itemsets that fall below the support threshold. What we envisage is dealing with Steps 3-7 together as discussed in Section 5.

The second phase of the query tree is to extract association rules from the frequent item sets generated in the first phase. Step 8 shows two copies of large itemsets (A and B), which are then joined on the condition that the itemset of a tuple in A is a subset of the itemset of a tuple in B (Step 9). The attributes are renamed to identify the smaller itemset as body (BD) of potential rules and its corresponding larger itemset as superset (sp). In step 11, the confidence values for candidate association rules are computed and then rules that have *confidence* greater than the *confidence threshold* are selected in Step 12. In Step 13, the head (HD) of the rules is obtained by applying the set difference operator, and the result consisting of the body (BD), head (HD), support and confidence are projected for all valid rules.

Figure 3 shows the intermediate results obtained by evaluation of the query tree for the sample data of Figure 1a. For convenience, we have omitted the commas and set braces from set values in Figure 3. For example, an itemset HJ denotes {H, J}. The intermediate results of steps 1-13 of the query tree are shown along with the corresponding algebraic expression for each step. As the sample input table and the final output are given in Figure 1, they are not repeated in Figure 3.

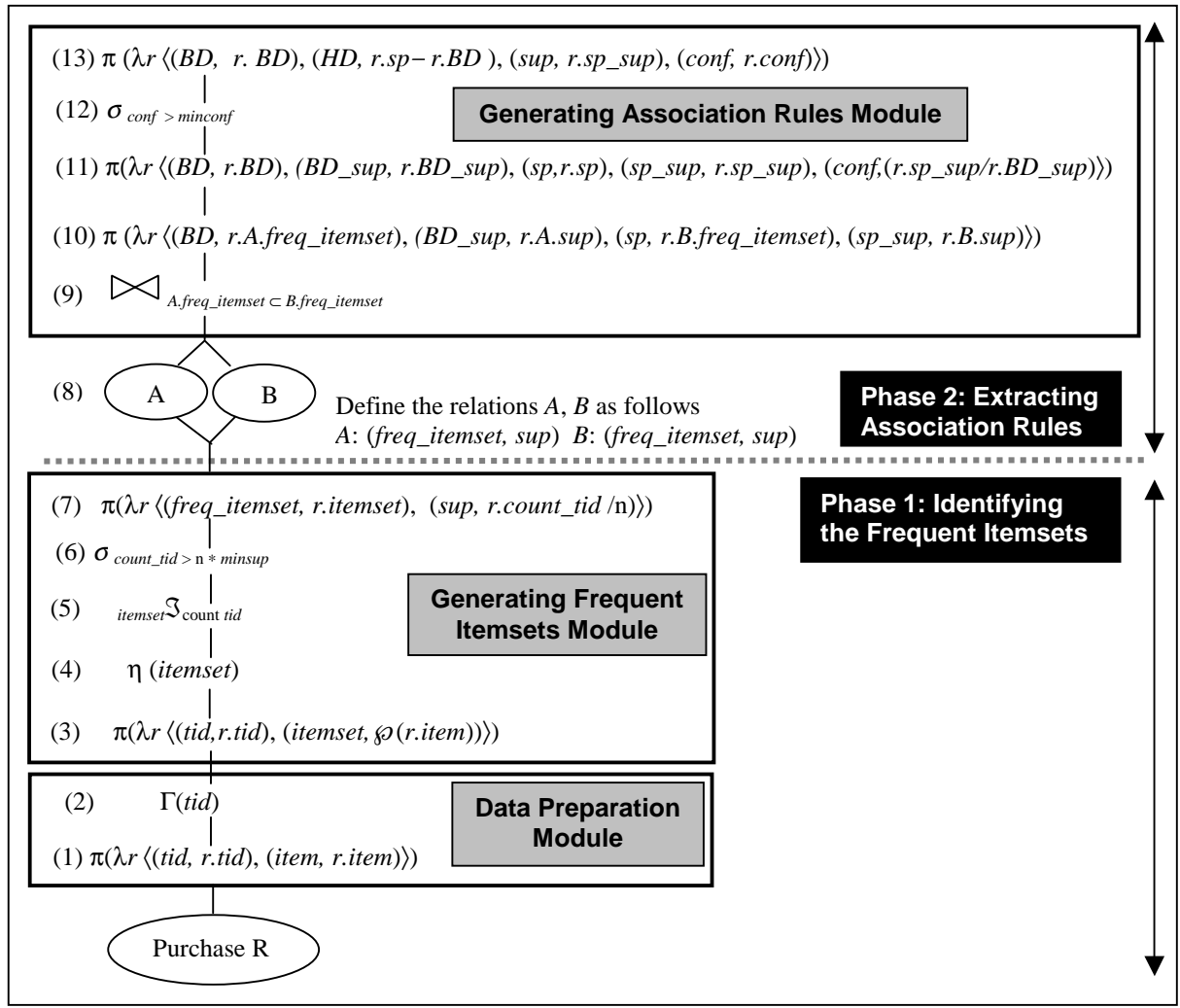

**Figure 2**: The Query Tree for Discovering Association Rules

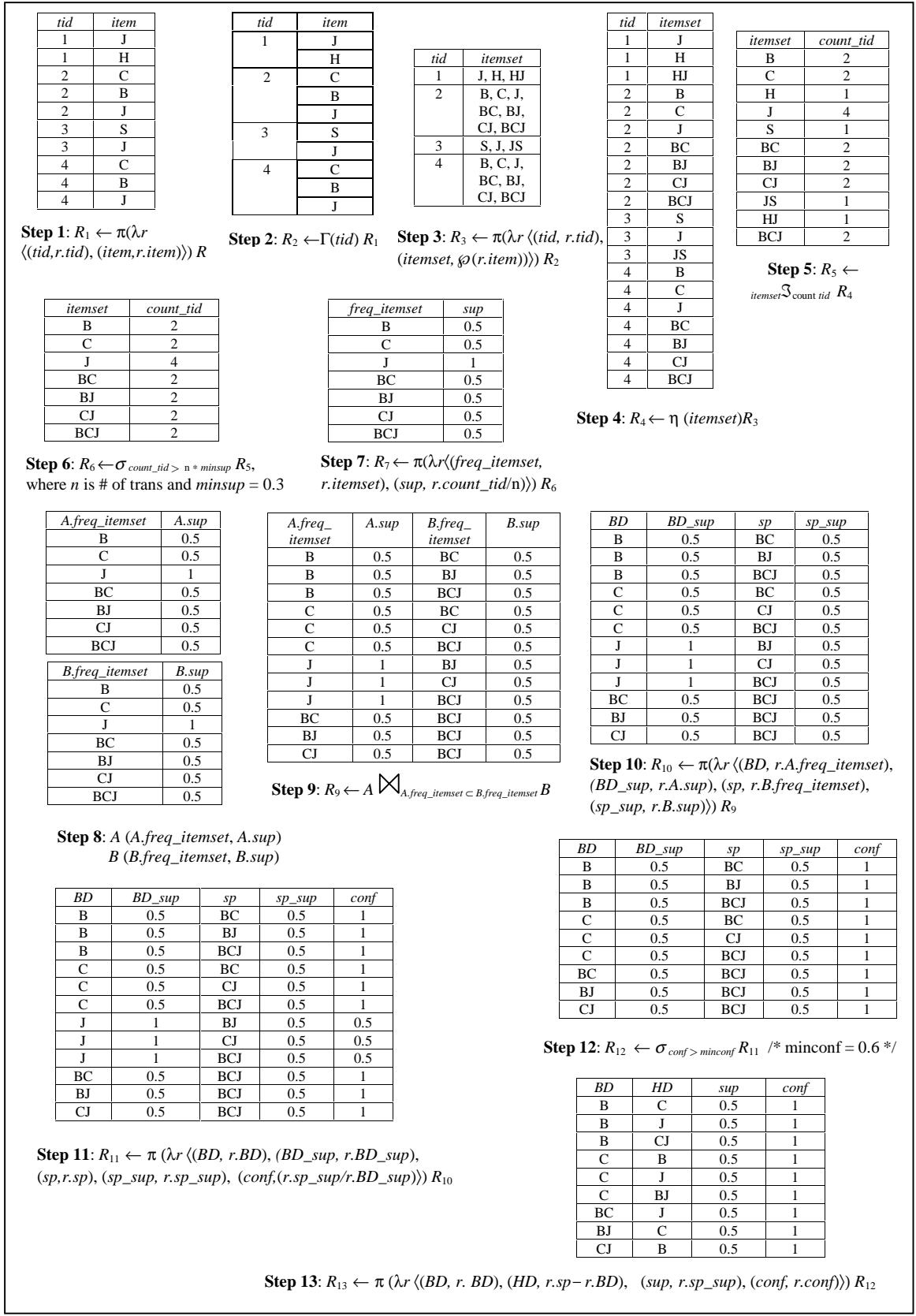

**Figure 3**: The Intermediate Results of Query Tree Evaluation

## 3.2 Modularization

In representing the different variations of association rule queries, we noticed that some sequences of steps in the query trees are repeated for the different variants. This led to the modularization of repeating sequences of operators by treating each sequence as a module. In Figure 2, we have identified three modules: Data Preparation Module, Generating Frequent Itemsets Module, and Generating Association Rules Module. The use of modules further simplifies the query tree. More importantly, it facilitates assigning lower level algorithms to the query tree in the optimization process as discussed in Section 5. However, a module in this context is not a rigid structure but only a sequential grouping of algebraic operators.

We illustrate the use of modules by an example. The query tree of Figure 4 represents a simple variant of association rules discovery. It corresponds to the following MINE RULE query of Meo et al. (Meo, Psaila & Ceri 1998a):

```
MINE RULE SimpleAssociations AS
SELECT DISTINCT 1..n item AS BODY, 1..1 item AS HEAD, SUPPORT, CONFIDENCE
FROM Purchase
GROUP BY transaction
       HAVING COUNT(*) <= 6
EXTRACTING RULES WITH SUPPORT: 0.1, CONFIDENCE: 0.2
```

The query specifies that only transactions of not more than 6 items need to be considered and the head of each rule should contain only one item. The query tree in Figure 4 uses the modules identified in Figure 2. After the data preparation module in Step 1, transactions with more than 6 items are eliminated from the input by Steps 2 and 3. The rules that have only one item in the HEAD are selected by means of Steps 7 and 8 of the query tree.

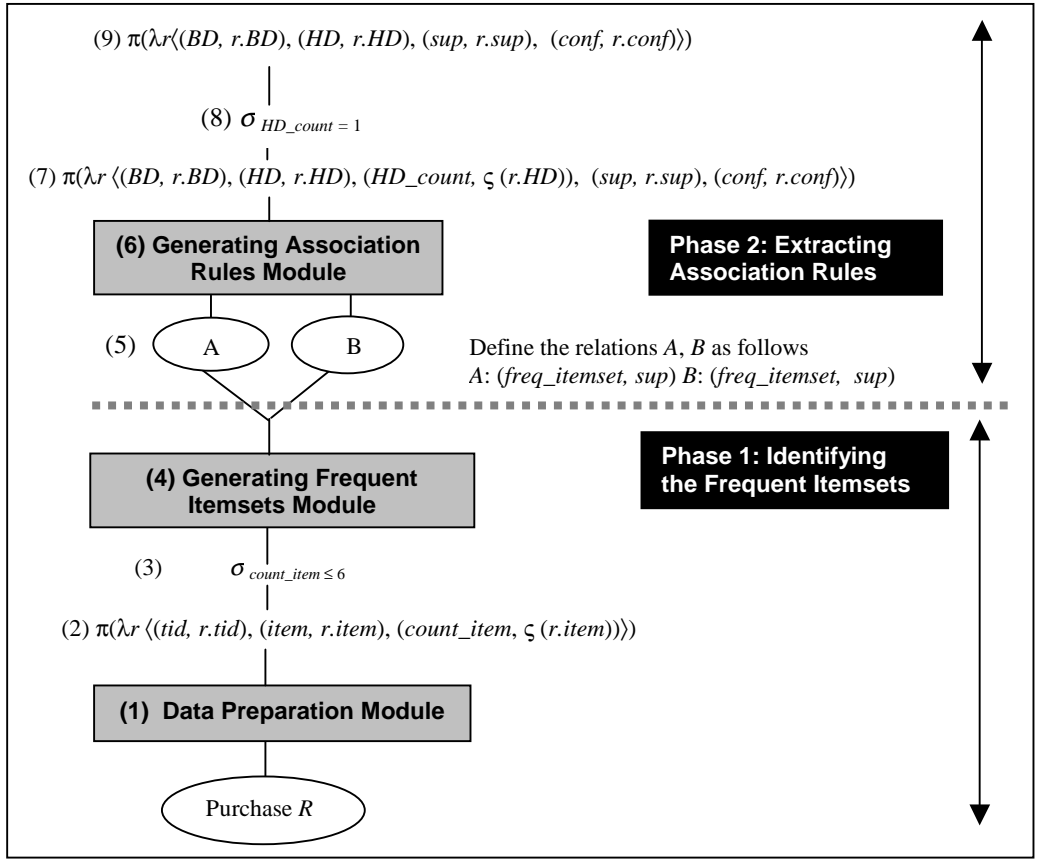

**Figure 4**: The Query Tree for a Variant of MINE RULE

# 4. Facilitating User Control in Association Rule Mining

Running a classical algorithm for mining association rules could take hours to complete. Users cannot stop or pause the process before it is completed and usually they get a huge list of rules. In order to deal with lack of user exploration and guidance in the data mining process, Ng et al. (Ng et al. 1998) proposed a 2-phase architecture that would allow for user interaction. They also used the concept of Constrained Association Queries (CAQ) to limit the computation to a subset of rules of interest to the user. In Section 4.1, we discuss how user control on the mining process is facilitated in our system. The representation of CAQs is discussed in Section 4.2.

## 4.1 Support of User Exploration and Control

The use of algebra for internal representation of association rule queries will facilitate user exploration and control of data mining systems. The algebraic query tree can be viewed as an abstract procedural algorithm and each node of the tree denotes a process step in the computation. It is possible to insert breakpoints for user interaction between any two nodes along the computation path of the tree. The user may be allowed to alter parameters or view intermediate results at these break points. They can also stop or pause further computation. Thus, the query tree provides a more flexible abstraction than conventional algorithms for supporting user interaction and control.

The user can change the support and confidence parameters during their interaction and thereby guide the mining process. For example, users can pause the process and change the value of *support threshold* before the computation reaches Step 6 in the query tree of Figure 2. Similarly, the *confidence threshold* can be modified before reaching Step 12. However, when modules are introduced into the query tree as discussed in Section 3.2, the break points within the module has to be mapped to the low-level algorithm that is chosen to implement the module by the optimization process. This is one of the issues being addressed in the development of a prototype of our optimizer.

## 4.2 Representing Constrained Association Queries

A *constrained association query* (CAQ) involves specifications of certain constraints on the antecedent and consequent of the rules to be mined. It is defined to be a query of the form: $\{(S_1, S_2) | C\}$, where $C$ is a conjunction of constraints on the set variables $S_1$ and $S_2$. Here we use the syntax of Ng et al. (Ng et al. 1998), but interpret $S_1$ and $S_2$ as the body and head of association rules instead of just any pair of frequent item sets.

Consider the following user query:

$\{(X_1, X_2) | X_1 \subset itemset$ and $X_2 \subset itemset$ and count $(X_1) = 2$ and count $(X_2) = 2\}$.

It is to find all pairs of itemsets having a cardinality of two. Assume a *minimum support* of 10% and a *minimum confidence* of 20%. As mentioned above, we consider $X_1$ as the body and $X_2$ as the head of the rules to be mined. For the sample data in the NewPurchase table of Figure 5a, the desired output of this query is shown in Figure 5b.

The query tree for this CAQ is shown in Figure 6. It is a modification of the query tree of Figure 2. By using the three modules identified in Figure 2, we are able to focus on the additional steps needed to represent the query constraints. Since the cardinality of $X_1$ and $X_2$ are equal to two, transactions with less than two items are eliminated in Step 3. We apply this constraint in Step 3 of the query tree after counting the items of each transaction in Step 2. Steps 5, 6 and 7 are used to filter the frequent item sets generated in Phase 1, so that the body and head of the rules will have the specified cardinalities.

To visualize the evaluation of this query tree, we provide the intermediate results for the sample data in Figure 7. The results of Steps 1-7 along with the corresponding algebraic expressions are shown. The input and output of the CAQ are given Figure 5, so they are not repeated in this figure.

| tid. | cust | item | date | price | qty | type |
|---|---|---|---|---|---|---|
| 1 | C1 | Scanner (S) | 17/06/2001 | 140 | 1 | peripheral |
| 1 | C1 | Hannibal (H) | 17/06/2001 | 40 | 1 | movie |
| 2 | C2 | Hannibal | 18/06/2001 | 40 | 2 | movie |
| 2 | C2 | CD-RW Driver (C) | 18/06/2001 | 250 | 2 | peripheral |
| 2 | C2 | Batman Returns (B) | 18/06/2001 | 35 | 1 | movie |
| 2 | C2 | Joystick (J) | 18/06/2001 | 30 | 1 | peripheral |
| 3 | C3 | Joystick | 19/06/2001 | 30 | 1 | peripheral |
| 4 | C4 | Batman Returns | 20/06/2001 | 35 | 1 | movie |
| 4 | C4 | CD-RW Driver | 20/06/2001 | 250 | 3 | peripheral |
| 4 | C4 | Joystick | 20/06/2001 | 30 | 2 | peripheral |

| BODY (BD) | HEAD (HD) | sup | conf |
|---|---|---|---|
| BC | HJ | 0.25 | 0.5 |
| BH | CJ | 0.25 | 1.0 |
| BJ | CH | 0.25 | 1.0 |
| CH | BJ | 0.25 | 1.0 |
| CJ | BH | 0.25 | 0.5 |
| HJ | BC | 0.25 | 1.0 |

(a) New Purchase Table for Department Store X  (b) Desired Output

**Figure 5**: Sample input data and desired output of a CAQ

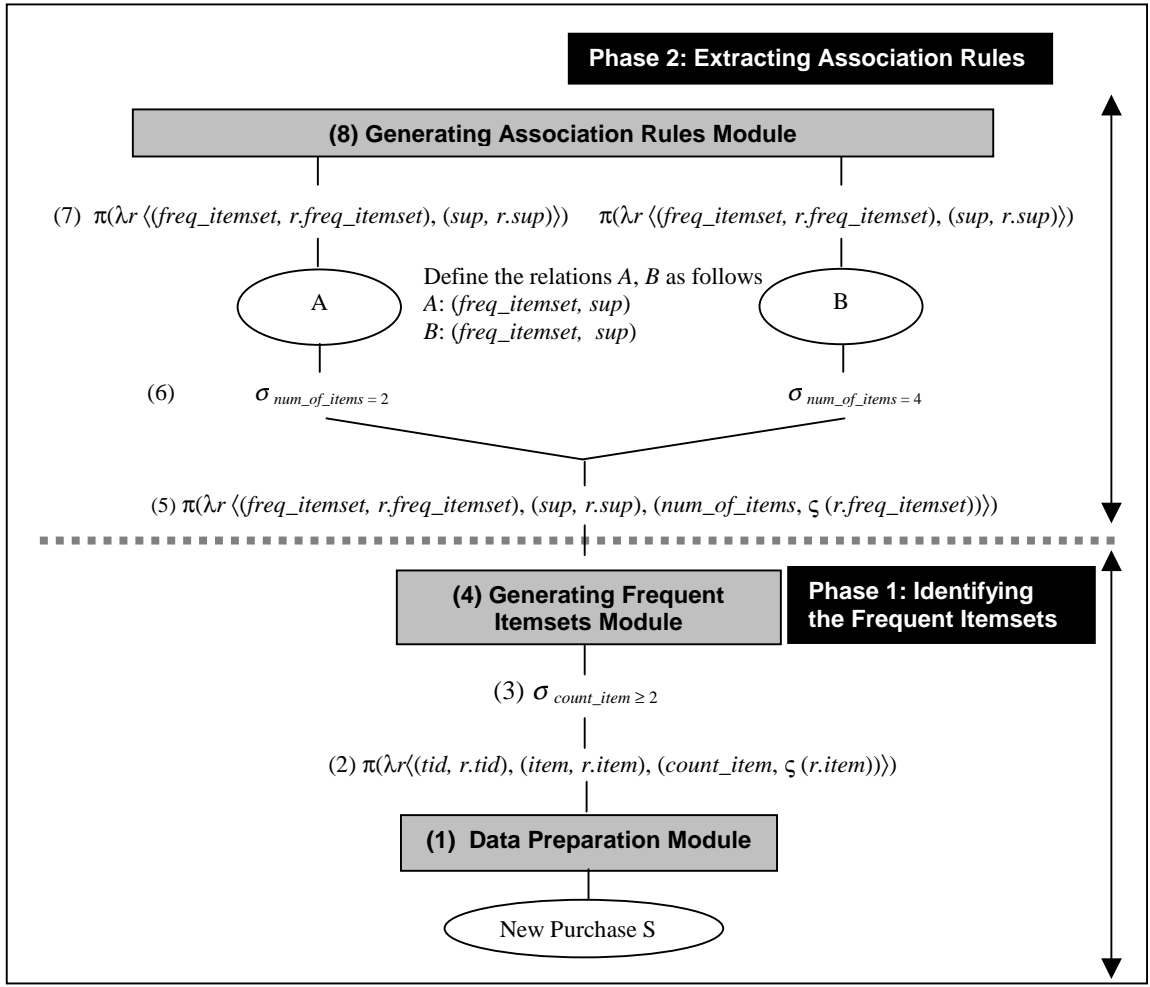

**Figure 6**: The Query Tree Representation of the Algebra with CAQ

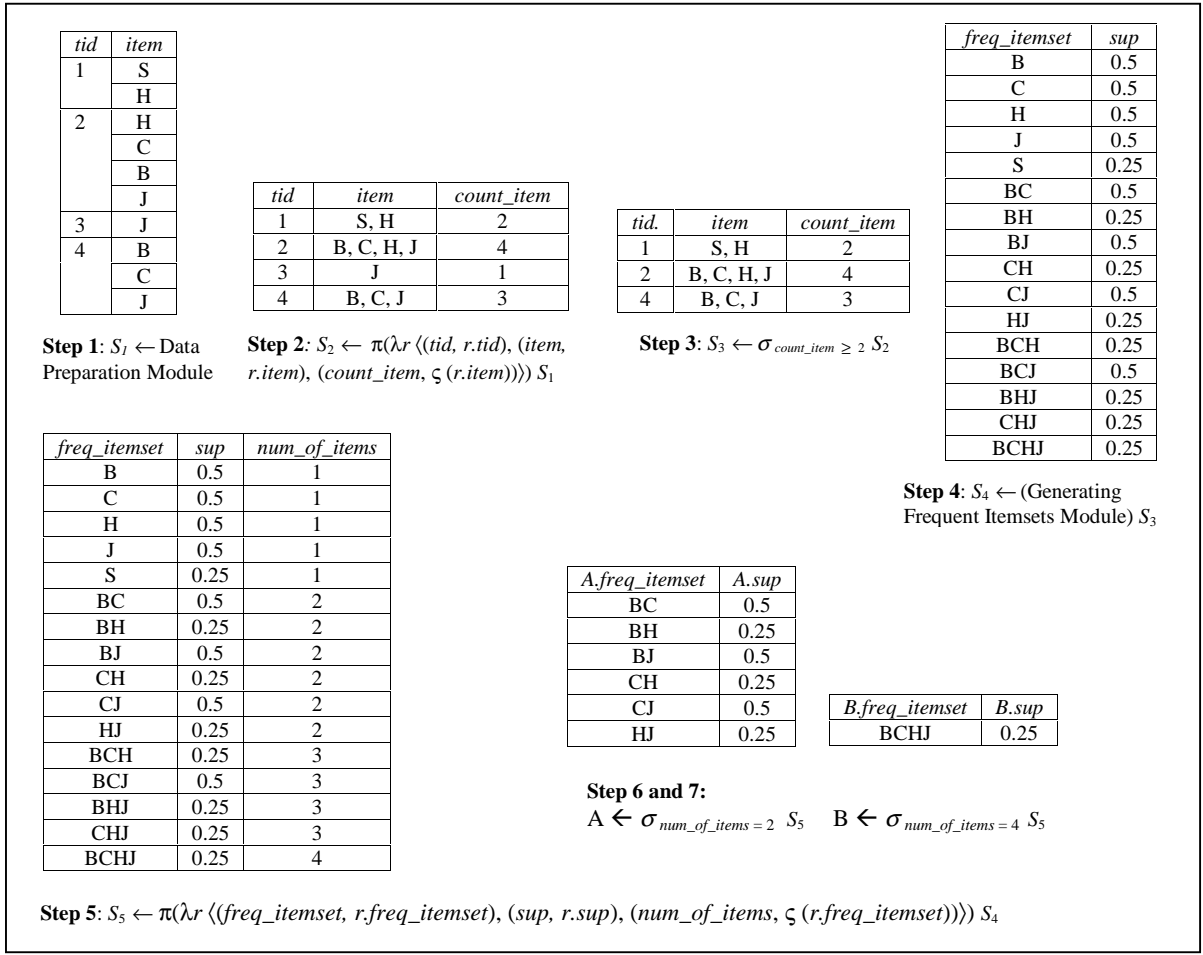

**Figure 7**: The Intermediate Results of the Query Tree for the CAQ

# 5. Optimization of Database and Data Mining Queries

Three main approaches have been followed in building data mining systems: tightly coupled with a DBMS, loosely coupled, and not using a database at all (Nestorov & Tsur 1999). In the tightly coupled systems, most of the data processing is performed on the database, taking advantage of facilities provided by the database system. In this paper, we propose a seamless integration of data mining and database systems as an alternative to the existing methods.

Our concept of integration is based on a common set of algebraic operators to express both database queries and data mining tasks. The optimizer for such a system can deal with both conventional database queries and data mining requests.

In this Section, we describe the framework for the common optimizer, and consider the optimization of data mining queries within this framework. As shown in Figure 8, the framework consists of seven stages that are similar to those of relational query optimizers. The different stages are described below.

**Stage 1: Accept Query**

The user submits a query written in a suitable query language. The query is parsed and checked for correct syntax. The user queries can be submitted in any of the available languages.

As mentioned in Section 1, several researchers have proposed query language extensions or language primitives for data mining such as DMQL, MINE, and MINE RULE. In our system, we use SQL extended with the features of MINE RULE and CAQs as our query specification language. However, we do not discuss the query language in this paper.

**Stage 2: Translation into Algebra**

The parsed user queries are translated into the extended nested algebra presented in Section 2. As the extended algebra is a super set of nested algebras needed for expressing database queries, it is suitable for representing both database and data mining queries. The query trees used for internal representation of association rule queries in our optimizer were described in Sections 3 and 4. The capability of the algebra to express different forms of association rule queries was also covered. The frequently used sequences of algebra operators are treated as modules as we discussed in Section 3.2.

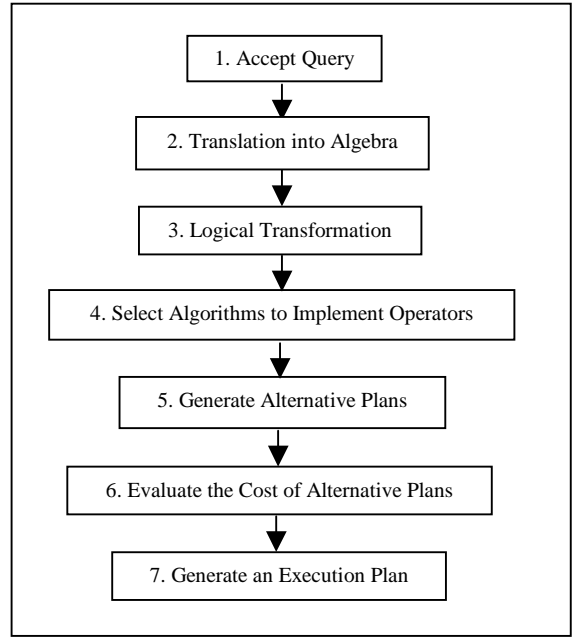

**Figure 8**: The Optimizer Framework

**Stage 3: Logical Transformation**

As in traditional query optimization, the objective of this stage is to reduce the size of intermediate results by using transformations that are independent of data values. The query tree is transformed by applying rules such as the SELECT, PROJECT, JOIN rule (Chaudhuri 1998b), pull-up transformation for group-by (Chaudhuri & Shim 1996), etc. As these transformations are well known, we do not describe them here.

We are currently developing rules that are specific to constrained association rule queries. These rules are intended to move the constraints of CAQs down in the query tree so that the cost of computation would be minimized. The different types of constraints that can appear in CAQs are considered in developing these transformation rules.

Another issue of significance is how to deal with the powerset operator. The use of a constrained powerset operator that combines steps of powerset generation with pruning of less frequent itemsets is being studied. It is somewhat similar to the concept of JOIN that combines the formation of Cartesian product with selection. However, the practical problem of performing powersets is overcome by the concept of modules as discussed below.

**Stage 4: Select Algorithms to Implement Operators**

There are several possible low-level algorithms to implement each operator. As an example, the alternatives for SELECT operator may include file scan, binary search, using a B+ tree index or a hash index. The choice will depend on several factors such as the existing indexes, size of buffer pool, sizes of the relations, sort orders, buffer replacement policy, etc. The main goal of this stage is to identify several ways to execute each operator in the query tree.

Apart from simplifying the query trees, the use of modules also helps in choosing alternative low-level algorithms. As in the case of individual algebra operators, each module can have several possible algorithms to implement it. For example, different algorithms for frequent item set generation in the literature such as Apriori, Hashing, Partition, DIC, OCD, and SETM, are being adapted for use as alternative low-level procedures in this stage. We are able to overcome the cost of computing the powersets by adopting this strategy.

As mentioned earlier, we are currently studying a constrained powerset operator for the algebra, that would combine powerset generation with pruning of small itemsets. However, at the lower level of implementation, it would not be significantly different from the existing procedures based on the Apriori algorithm.

**Stage 5: Generate Alternative Plans**

Alternative plans are generated from combinations of alternative algorithms for the various operators. In a plan, every node of the query tree is thus associated with an algorithm to execute the corresponding operator. The plans can include existing data mining algorithms in the literature as alternative low-level procedures, particularly for frequent itemset generation as discussed before.

**Stage 6: Evaluate the Cost of Alternative Plans**

Suitable formulae are used to compute the cost of execution of each algorithm so that the cost of alternative query plans can be estimated. The cost of each alternative plan is estimated by considering all its operations. The plan with the minimum cost should be considered as the best plan.

**Stage 7: Generate an Execution Plan**

The plan that has the lowest cost is chosen as the execution plan.

In this paper, we have focused on the internal representation scheme for the optimizer as well as some aspects of logical transformation of query trees and choice of low-level procedures. Further details of the optimizer design and implementation will be covered in a subsequent paper.

# 6. Conclusions

We presented a scheme for seamless integration of data mining with database systems. It is based on a common query optimizer for data mining and database systems, using an extended nested relational algebra for internal representation of association rule and other queries in the optimizer. The expressiveness of the algebra is shown by suitable examples of typical queries including a constrained association query. The algebraic expressions are presented as query trees. We discussed modularization of common sequences of algebra operations to simplify the query tree for data mining. The concept of modules also facilitates the generation of alternative query plans for query optimization using existing algorithms in the literature as low-level procedures.

The algebraic formulation of an association rule query can be viewed as an abstract procedural algorithm. The users get more control over the mining process by being able to stop or pause the computations along particular branches of the query tree. This is achieved by providing optional break points between nodes of the query tree.

As far as we know, our approach is distinct from all existing systems discussed in the literature. The use of a common algebra for integrating data mining with database systems has not been proposed before. Though the algebraic operators we use are not new and the query optimization scheme is the same as for relational databases, what we describe is distinctly different from previous approaches to integration. It is also conceptually clearer and simpler than most other proposals.

This research is part of a larger project for integrating data mining and database management. In this paper, we have focused on the query optimizer framework, the internal representation scheme for database and data mining queries, and some significant aspects of the early stages in the optimization process. The development of a prototype is currently underway. Further details of the optimizer design and implementation will be described in a subsequent paper.